# Phase-sensitive optical pulse characterization on a chip via Spectral Phase Interferometry for Direct Electric-Field Reconstruction (SPIDER)

Alessia Pasquazi[1], Marco Peccianti[1], Yongwoo Park[1], Brent E. Little[2], Sai T. Chu[2-3], Roberto Morandotti[1], José Azaña[1], and David J. Moss[4] **

[1] Ultrafast Optical Processing, INRS-EMT, Université du Québec,
1650 Blv. L. Boulet, Varennes, Québec J3X 1S2 Canada
[2] Infinera Corp. 9020 Junction drive Annapolis, Maryland, 94089. USA
[3] Department of Physics and Materials Science, City University of Hong Kong
[4] CUDOS, School of Physics, University of Sydney, New South Wales 2006, Australia
** D.J. Moss Present Address: School of Electrical and Computer Engineering, RMIT University, Melbourne, Victoria, Australia 3001
email alessia.pasquazi@gmail.com

The recent introduction[1-3] of coherent optical communications has created a compelling need for ultra-fast phase-sensitive measurement techniques operating at milliwatt peak power levels and in time scales ranging from sub-picoseconds to nanoseconds. Previous reports of ultrafast optical signal measurements [4-7] in integrated platforms[8-10] include time-lens temporal imaging[5] on a silicon chip[8,9] and waveguide-based Frequency-Resolved Optical Gating (FROG)[4,6,10]. Time-lens imaging is phase insensitive while waveguide-based FROG methods require the integration of long tuneable delay lines -still an unsolved challenge. Here, we report a device capable of characterizing both the amplitude and phase of ultrafast optical pulses with the aid of a synchronized incoherently-related clock pulse. It is based on a novel variation of Spectral Phase Interferometry for Direct Electric-Field Reconstruction (SPIDER)[4,7] that exploits degenerate four-wave-mixing (FWM) in a CMOS

**compatible chip. We measure pulses with <100mW peak power, a frequency bandwidth >1THz, and up to 100ps pulsewidths, yielding a time-bandwidth product (TBP)>100[11].**

The accurate measurement of the amplitude and phase profiles of ultrafast optical signals is critical for a wide range of applications[4], ranging from metrology to coherent optical telecommunications[1-3]. Interferometry is broadly used in optics to measure the phase and amplitude of the electric field[4,12]. In holography, complete information on the spatial light distribution is obtained by recording the fringes resulting from the interference with a reference field. Likewise, the complex shape of a pulse under test (PUT) can be extracted by interfering it with a well-characterized reference pulse[4,12]. Spectral shearing interferometry uses a frequency detuned (*sheared*) replica of itself as a reference[4, 7,13]. From the rapid modulation in spectral intensity (from the spectral fringes), it is possible to unambiguously extract the PUT phase with a simple filtering procedure and algebraic concatenation[14, 15]. The spectral shear can be generated by a non-linear optical-parametric process, as in SPIDER[7,15], or by linear methods using synchronized electro-optic phase modulators[16,17]. Proposed in 1998 by Iaconis and Wamsley[7,15], SPIDER and its many variants have been recognized as an exceptional tool for characterizing pulses[4,13-24]. Its success is due to its intrinsically ultrafast[18] and single-shot[19] nature, as well as to its simple, direct and robust phase retrieval procedure[14,15]. Its different implementations enable pulse characterization over a very broad wavelength range, from the near infrared to the ultraviolet[4,13,23].

In the classical implementation of SPIDER, two replicas of the PUT are delayed and mixed with a strongly chirped pump pulse via Three Wave Mixing (TWM). The two replicas interact with different portions of the chirped pump pulse and experience mixing with different pump frequencies. The generated idler results in the superposition of two frequency-shifted replicas of the PUT. The classical SPIDER derives the pump from the PUT[4,7,15], making the technique intrinsically self-referencing. Alternatively, the use of a well-characterized pump reference pulse can improve the accuracy and reliability of the method[20], and this approach is commonly named X-SPIDER[4,21,23]. Typical SPIDER and X-SPIDER methods are designed for the characterization of optical pulses shorter than 100fs[4] and peak powers exceeding 10kW, and hence are not ideally suited to telecommunications. For these applications, where larger sensitivities and TPBs are required, approaches based on highly performing Lithium Niobate bulk electro-optic phase modulators have proven suitable for milliwatt peak powers and pulsewidths from 0.5ps to 100ps[16,17]. A SPIDER device capable of measuring optical pulses with these characteristics directly on a chip, particularly if the integration platform were compatible with electronic technology such as CMOS, would represent a fundamental advance not only for applications to optical telecommunications but also for many other areas such as computing microchips[3]. However, both TWM and linear electro-optic modulation require a second-order nonlinearity that significantly restricts the range of suitable materials, making the implementation of classical SPIDER devices in a CMOS compatible platform extremely challenging.

In this paper, we demonstrate phase sensitive ultrafast optical pulse measurements using a novel X-SPIDER approach based on degenerate FWM[24]. This $\chi^{(3)}$ (third-order) nonlinear process occurs in centro-symmetric materials such as glass and silicon – the latter being the basis of CMOS integration platforms. Our approach maintains the relative roles of the signal, the (external) pump and the idler that exist in the classical TWM implementation. The nonlinear optical component in our work is a 45-cm long spiral waveguide implemented in a high index, CMOS compatible, doped silica glass platform[25-28] (see Methods for details). The high nonlinear response exhibited by these waveguides[27,28] allows efficient wavelength conversion[25,26,28] throughout its anomalous dispersion regime, spanning from 1300nm to 1600nm[26].

In addition we generalize the phase-recovery algorithm previously used for X-SPIDER techniques in order to extend its operating time window. The classic SPIDER retrieval algorithm targets the spectral phase-recovery of the PUT from the measured spectral interferogram. Therefore, the two PUT replicas must be short enough compared to the stretched pump in order to interact with an approximately monochromatic region of the pump, so that the two resulting idlers are not spectrally distorted. This condition fails for a highly chirped PUT, since nonlinear mixing occurs over a range of pump frequencies. However, the resulting spectral interferogram still contains complete information of the PUT phase, and so in principle it is possible to extract the correct phase by applying appropriate corrections to the phase extraction procedure, as previously demonstrated for Gaussian pulses in a self-referenced setup[22]. Here, we use the fact that our X-SPIDER approach actually measures *the Fresnel integral of the PUT* (see Methods), dispersed by the chirp imposed on the idler via the FWM interaction, equivalent to half of the pump chirp. This integral can be inverted to retrieve the PUT, once the delay and pump chirp are known; this knowledge is also required by the classic X-SPIDER algorithm. Our algorithm works well as long as the interference fringes are recorded in the interferogram and can be applied to stretched pulses with a much larger TBP than the standard SPIDER algorithm, which is typically limited to TBPs <10[22].

If combined with this new algorithm, our FWM X-SPIDER device exhibits performance suited for ultrafast coherent optical communications. When supplied with an incoherently-related optical reference clock (i.e. a pulse at a different wavelength, often available in optical communications systems) the device performs phase-sensitive measurements of optical pulses with bandwidths >1THz in the C and L bands, at peak powers less than 100mW, and over time windows of up to ~100ps, yielding a TBP > 100. This is among the highest TPB performance, with comparable sensitivity, reported to date for any sheared spectral interferometry implementation, including linear designs with state of the art bulk electro-optic modulators[16,17].

The experimental setup is shown in Figure 1(b), and consists of an all-fibre configuration to prepare a generic PUT and a well-calibrated pump pulse to input into the FWM X-SPIDER device. We prepared a PUT through self-phase-modulation induced spectral broadening, with a bandwidth of ~1THz (full-width at -10dB of 1.15THz, 1.05THz and 1.1THz for the pulses in Figure 2 (a), (c), (d) respectively). The pulse energy at the input of the chip was < 5pJ and we applied different spools of standard single-mode fibre (SMF) to control the chirp, stretching the pulse up to 100ps, corresponding to ~70mW peak power. In the low-chirp regime, the spectra of the two replicas of the idler were sheared copies of the signal spectrum (with low and high frequency components switched in position) (Figure 2 (a,b)). For high chirp the two idler replicas showed significant spectral distortion when compared with the signal (Figure 2 (c,d,e,f)). The delay was accurately inferred via the linear interference of the superimposed signal replicas appearing in the measured spectrum. This approach minimizes the measurement error on the delay, a fundamental parameter for SPIDER techniques[29].

Figure 3 shows the full set of results for our FWM X-SPIDER measurements using both the classic and extended (Fresnel) algorithms, along with measurements using a SHG-based FROG technique. As expected, for low TBP pulses (i.e., the short-pulse regime) our FWM X-SPIDER approach yielded the same solution when using either phase-recovery algorithms (Figure 3, Panel (I), (e,f)) and both of these agreed well with the experimental FROG spectrogram (Figure 3, Panel (I), (g)). For large TBP pulses (highly chirped, long pulsewidths) the standard phase-recovery algorithm yielded a pulse spectrogram profile that deviated significantly from the experimental FROG trace. On the other hand, the spectrogram obtained using the extended (Fresnel) phase-recovery algorithm agreed very well with the FROG trace, thus confirming the extended range of TBP operation for our FWM X-SPIDER device.

The accuracy of our phase extraction process was estimated by measuring the dispersion introduced on Gaussian pulses by well-calibrated SMF spools (Figure 4). We stretched a transform-limited Gaussian pulse with a FWHM bandwidth of ~0.53THz up to ~70ps, i.e. more than 100 times its transform-limited length, by using spool lengths from 1.09 km to 1.15 km. The signal pulse energy coupled into the chip in this configuration was < 10pJ, corresponding to a peak power of ~100mW. In all cases the expected quadratic phase curvature was retrieved (Figure 4 (a-f)) with excellent agreement - the phase difference induced by the 10m increments of fibre length is estimated with an error < 5%. Therefore our method can accurately resolve changes in dispersion as small as the equivalent of 50cm of SMF with a total dispersion of more than 1km of SFM (Figure 4 (g)).

The SPIDER device demonstrated here was enabled by the high linear and nonlinear optical performance of our platform; however we expect that it should be also achievable in silicon, since the required nonlinear FWM conversion efficiency is comparable to that needed for time lensing demonstrated in silicon nanowires[8]. Given an externally supplied pump pulse, full integration for processing the PUT on chip can be readily achieved by integrating a fixed delay interferometer.

In conclusion, we report the demonstration of an optical oscilloscope capable of measuring both the phase and amplitude of THz-bandwidth optical pulses, with a time bandwidth product exceeding 100, and peak powers <100mW. This device, based on a CMOS compatible chip, uses a novel X-SPIDER technique that relies on FWM rather than conventional TWM. We achieve a sensitivity to total dispersion equivalent to a length of standard SMF as small as 50cm. We believe this work represents a key milestone in achieving full characterization of complex ultrafast optical waveforms on a chip.

**Methods**

**Theoretical Approach and Reconstruction Algorithm.** Degenerate FWM in the spiral waveguide involves two replicas of the PUT (characterized by a complex-field envelope $e(t) = \int_{-\infty}^{\infty} E(\omega)\exp(-i\omega t)d\omega$ and $E(\omega) = |E(\omega)|\exp(i\varphi_E(\omega))$, in the temporal and spectral domains, respectively) delayed by $\Delta t$ and overlapped with a pump pulse $p(t)$. The pump pulse is temporally stretched by the propagation through a predominantly first-order dispersive element, resulting in a temporal phase curvature $\phi_P = \beta_2 L$ (with $L$ and $\beta_2$ being the length and group velocity dispersion of the dispersive element, respectively). The pump pulse envelope $|p(t)|$ is assumed to be constant over the entire duration of the two delayed PUT pulses: $p(t) \approx \exp\left(\frac{it^2}{2\phi_P}\right)$, a condition that can be easily fulfilled by dispersing many classes of transform-limited pulses, e.g. spectrally bell-shaped or flat-top pulses. The idler $s(t)$ induced by the (non-depleted) degenerate FWM interaction is simply given by the following expression:

$$\overline{s(t)} \propto \left[e(t+\Delta t)+e(t)\right]\overline{p(t)^2} \approx \left[e(t+\Delta t)+e(t)\right]\exp\left(-\frac{it^2}{\phi_P}\right) \tag{1}$$

where the upper line represents the complex conjugate operation. Higher order interaction terms such as SPM, XPM, as well as secondary idler generation are neglected, consistent with the operational characteristic of our device. The formula for the degenerate FWM X-SPIDER is equivalent to the usual relation describing difference frequency generation in TWM X-SPIDERs[23], except that the effective chirp of the interaction is half of the pump chirp in this case. In the spectral domain we can write:

$$\overline{S(-\omega)} \propto \left[E(\omega)\exp(-i\omega\Delta t)+E(\omega)\right]*\exp\left(\frac{i\omega^2\phi_P}{4}\right) \tag{2}$$

Where the symbol * represents a convolution operation. We define $f_e(t)$ as the Fresnel integral of the PUT corresponding to its dispersion through half the length of the dispersive element, i.e.

$$f_e(t) = e(t)*\exp\left(-\frac{it^2}{\phi_P}\right) = \exp\left(-\frac{it^2}{\phi_P}\right)\int_{-\infty}^{\infty} e(x)\exp\left(-\frac{ix^2}{\phi_P}+\frac{i2xt}{\phi_P}\right)dx \tag{3}$$

It is thus possible to readily write the relation between the spectrum $F_e(\omega)$ (Fourier transform of the Fresnel integral in Eq. (3)) and the PUT spectrum - i.e. $E(\omega) \propto F_e(\omega)\exp\left(-\dfrac{i\omega^2\phi_P}{4}\right)$ - and then substitute this into Eq. (2). It is easy to verify that the following relationship holds:

$$\left[F_e(\omega)\exp(-i\omega\Delta t)\exp\left(-\dfrac{i\omega^2\phi_P}{4}\right)\right] * \left[\exp\left(\dfrac{i\omega^2\phi_P}{4}\right)\right]$$
$$= \exp\left(\dfrac{i\omega^2\phi_P}{4}\right)\int_{-\infty}^{\infty} F_e(s)\exp\left[-is\left(\dfrac{\omega\phi_P}{2}+\Delta t\right)\right]ds =$$
$$= \exp\left(\dfrac{i\omega^2\phi_P}{4}\right)f_e\left(\dfrac{\omega\phi_P}{2}+\Delta t\right) \tag{4}$$

Using the relationship in Eq. (4), the resulting idler energy spectrum, calculated from Eq. (2), can then be expressed as a function of the modulus $|f_e(t)|$ and phase $\varphi_{fe}(t)$ of the Fresnel integral:

$$|S(-\omega)|^2 \propto \left|f_e\left(\dfrac{\omega\phi_P}{2}+\Delta t\right)\right|^2 + \left|f_e\left(\dfrac{\omega\phi_P}{2}\right)\right|^2 +$$
$$+2\left|f_e\left(\dfrac{\omega\phi_P}{2}+\Delta t\right)\right|\left|f_e\left(\dfrac{\omega\phi_P}{2}\right)\right|\cos\left[\varphi_{fe}\left(\dfrac{\omega\phi_P}{2}+\Delta t\right)-\varphi_{fe}\left(\dfrac{\omega\phi_P}{2}\right)\right] \tag{5}$$

When the PUT duration is sufficiently short (e.g. when the PUT chirp is negligible compared to the (large) pump chirp), we can then carry out a Fraunhofer approximation of the Fresnel integral in Eq. (3):

$$f_e(t) = e(t) * \exp\left(-\dfrac{it^2}{\phi_P}\right) \approx E\left(\dfrac{2t}{\phi_P}\right)\exp\left(-\dfrac{it^2}{\phi_P}\right) \tag{6}$$

By introducing the approximation expressed by Eq. (6) into Eq. (5), one can easily infer that the interferogram results in the classical expression for the spectral-sheared interference pattern that is typically used for phase-recovery in standard SPIDER:

$$|S(-\omega)|^2 \propto |E(\omega+\Omega)|^2 + |E(\omega)|^2 + 2|E(\omega+\Omega)||E(\omega)|\cos\left(\varphi_E(\omega+\Omega) - \varphi_E(\omega) - \Delta t\omega - \frac{\Delta t\Omega}{2}\right)$$

(7)

with a spectral shear given by $\Omega = 2\Delta t / \phi_P$.

The standard SPIDER phase-reconstruction algorithm addresses the interferogram as expressed by the approximate relation (6), thus being valid only under the relatively limited conditions of the Fraunhofer approximation defined above (typically, low signal chirp). This approximation is strictly valid only when the following condition is satisfied:

$$\Delta\tau_e^2 << 2\pi\phi_P \approx \frac{\Delta\tau_P}{\Delta\nu_P} \qquad (8)$$

where $\Delta\tau_{e(P)}$ and $\Delta\nu_{e(P)}$ are the full-width temporal duration and frequency bandwidth for the PUT $e(t)$ and for the pump $p(t)$, respectively. Considering that the pump frequency bandwidth typically exceeds that of the PUT and $1/\Delta\nu_e < \Delta\tau_e/\pi$, it can be easily inferred from the inequality in Eq. (8) that the PUT time width must necessarily be much smaller than the pump time duration, $\Delta\tau_e << \Delta\tau_P$. This estimate clearly points out the limited pump time window applicable for the classical SPIDER algorithm.

In our more general algorithm, we use the exact interferogram pattern defined by relation (5) to extract the PUT spectral phase information. In particular, as long as the Fourier-filtering procedure that is conventionally used in sheared interferometry[14,15] can be applied, it is possible to extract the phase profile of the Fresnel integral $f_e(t)$ (instead of the direct PUT spectral phase profile) from the measured spectral interferogram (using the same procedure as for the classical SPIDER phase-reconstruction algorithm). In parallel with this, the amplitude of this function can be easily obtained from a separate measurement (i.e. by blocking one arm of the interferometer), as it is usually done in the standard SPIDER implementation. Once the Fresnel integral function is completely known (in amplitude and phase), it can be numerically inverted in order to find the complex field of the PUT. This approach is simple and accurate and requires the same experimental information as the classical X-SPIDER (only the pump dispersion $\phi_P$ and the delay $\Delta t$ must be accurately known). It significantly extends the working range of the instrument, enabling the measurement over a temporal window as large as the square of the pump pulse duration.

**Experiment**

Pump and signal pulses were obtained from a 17MHz mode-locked fibre source, providing tuneable pulses in the range 1530nm-1560nm. The chirped pump was prepared via spectral broadening of the pulses generated by a mode-locked fibre laser in a nonlinear fibre, followed by filtering with a detuned bandpass optical filter, and then temporally stretched via linear dispersion through a SMF fibre, finally being amplified with an EDFA.

The two replicas of the PUT are delayed of about 6ps. The delay was inferred via the linear interference of the superimposed signal replicas appearing in the measured spectrum, accurately measurable in the range ~1.5ps-400ps for our system.

The lengths of the SMF spool were measured by propagating a 250ps modulated optical pulse, detecting the transmitted pulse through a fast photodetector and a sampling scope, in order to finally determine its time-of-flight.

**Device**

The spiral waveguide was fabricated in the Hydex glass platform[27,28]. Films were deposited by plasma-enhanced chemical vapor deposition (PECVD)[25], and yielded low loss waveguides (at 1550nm) as deposited, without the requirement of high temperature annealing, thus making them ideal for a CMOS compatible fabrication process. Device patterning and fabrication were performed using photolithography and reactive ion etching to produce low sidewall roughness on the core layer, before over-coating with a silica glass upper cladding layer. The resulting waveguide possesses a high effective nonlinearity (220W$^{-1}$km$^{-1}$) due to a combination of tight mode confinement and high intrinsic $n_2$, as well as a small and anomalous dispersion[26], and has very low linear (<6dB/m) and negligible nonlinear optical losses (up to 25GW/cm$^2$)[25,26]. This platform has shown excellent FWM performance - both in terms of conversion efficiency and bandwidth - with no appreciable gain saturation[28].

**Acknowledgements**

This work was supported by the Australian Research Council (ARC) Centres of Excellence program, the Fonds Québécois de la Recherche sur la Nature et les Technologies (FQRNT), the Natural Sciences and Engineering Research Council of Canada (NSERC), NSERC Strategic Projects, and the INRS. M. P. acknowledges the support of the Marie Curie People project TOBIAS PIOF-GA-2008-221262. We are thankful to Matteo Clerici for enlightening discussions and technical help.

**Authors' contribution**

A.P. designed the experiment, the novel algorithm and analysed experimental data. A.P., Y. P. and M.P. ran the experiment, M.P. built and calibrated the FROG set-up for the experiment validation. J.A. contributed in the theoretical derivation. S.T.C. and B.E. designed and build the sample. J.A., R.M. and D.M. supervised the project. A.P., M.P, R.M., J.A. and D.M. wrote the manuscript.

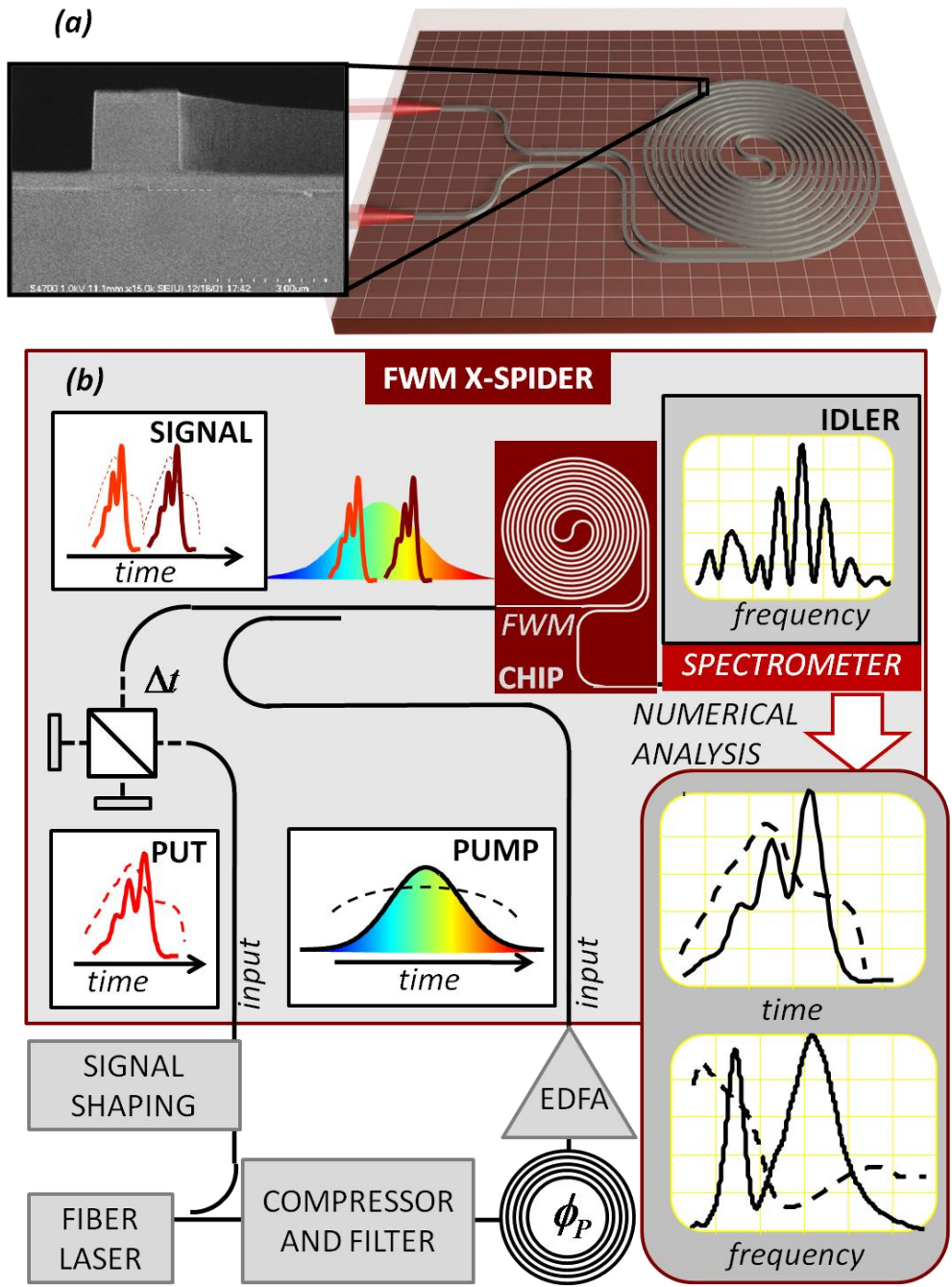

Figure 1: Device structure. (a) Diagram of the spiral waveguide. Inset: micrograph of the waveguide cross section. (b) X-SPIDER based on FWM: the PUT is split in two replicas and nonlinearly mixed with a highly chirped pump inside the chip (extracted from the same laser source in our set-up). The resulting output is captured with a spectrometer and numerically processed to extract the complete information (amplitude and phase) of the incident pulse. More details on the set-up are in the Methods Section.

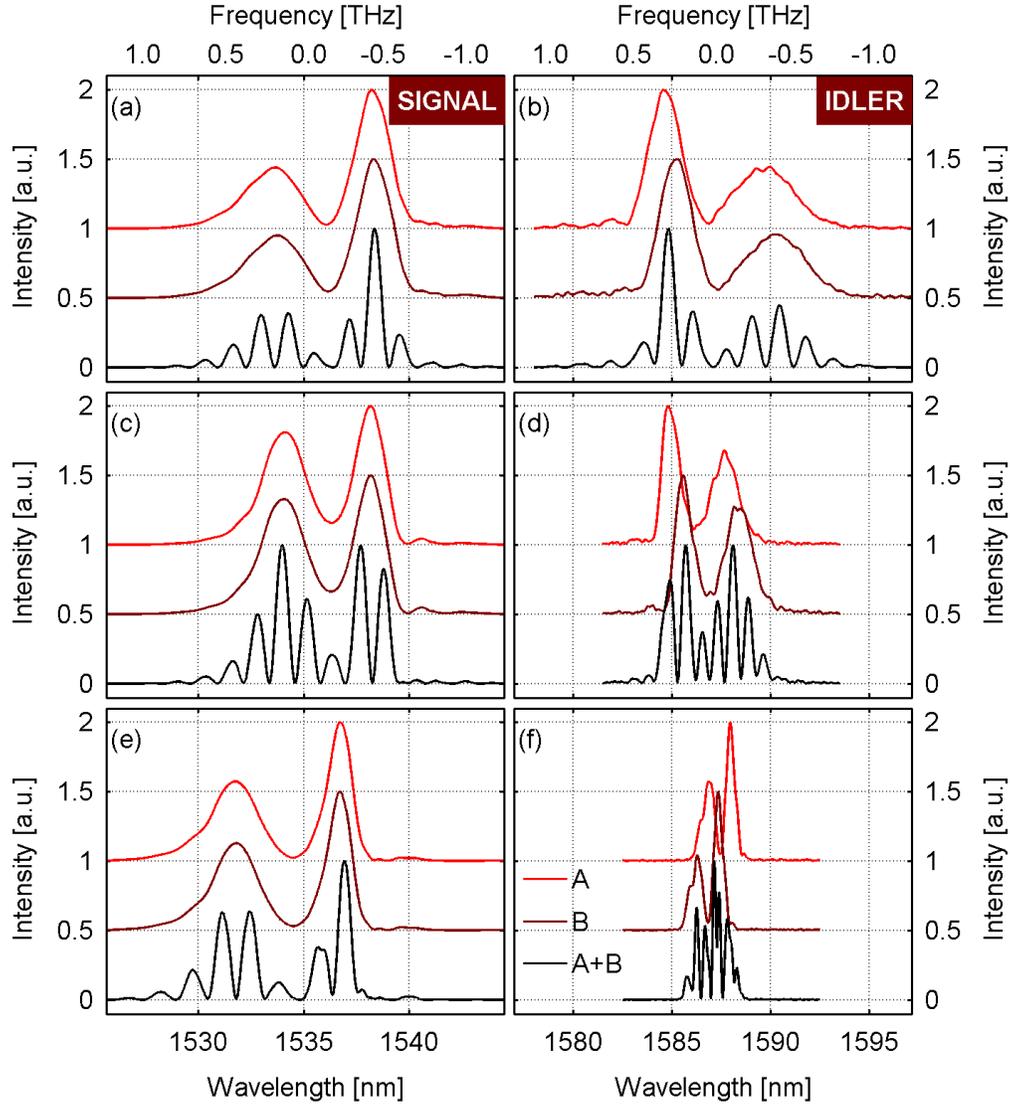

Figure 2: Resulting output spectra of the FWM-X-SPIDER device for a complex waveform of ~ 1THz bandwidth. Here the pump is a 10nm bandwidth (FWHM=1.2THz) pulse centred at 1565nm and stretched to 220ps, with a 800pJ energy, which results in a super-Gaussian (highly chirped) profile. A standard calibration technique was employed[15] to extract its phase curvature of ~20ps$^2$, resulting in an equivalent chirp for the nonlinear interaction of 10ps$^2$ (see Methods for details). (a-b), (c-d) and (e-f) display pulses with a full-bandwidth at -10dB of 1.15, 1.05 and 1.1 THz and a TBP of 5, 30 and 100, respectively. Spools of SMF fibre of 150m and 650m in length have been used to control the dispersion of the PUT in the cases (c-d) and (e-f), for a total dispersion of ~ 3.5ps$^2$ and 14ps$^2$, respectively. Spectra of the two PUT delayed replicas alone (A and B) and superimposed (A+B), for the signal (a, c and e) and for the idler (b, d and f), the latter generated through the nonlinear FWM process taking place inside the waveguide. Both the idler and signal are simultaneously measured using the same spectrometer.

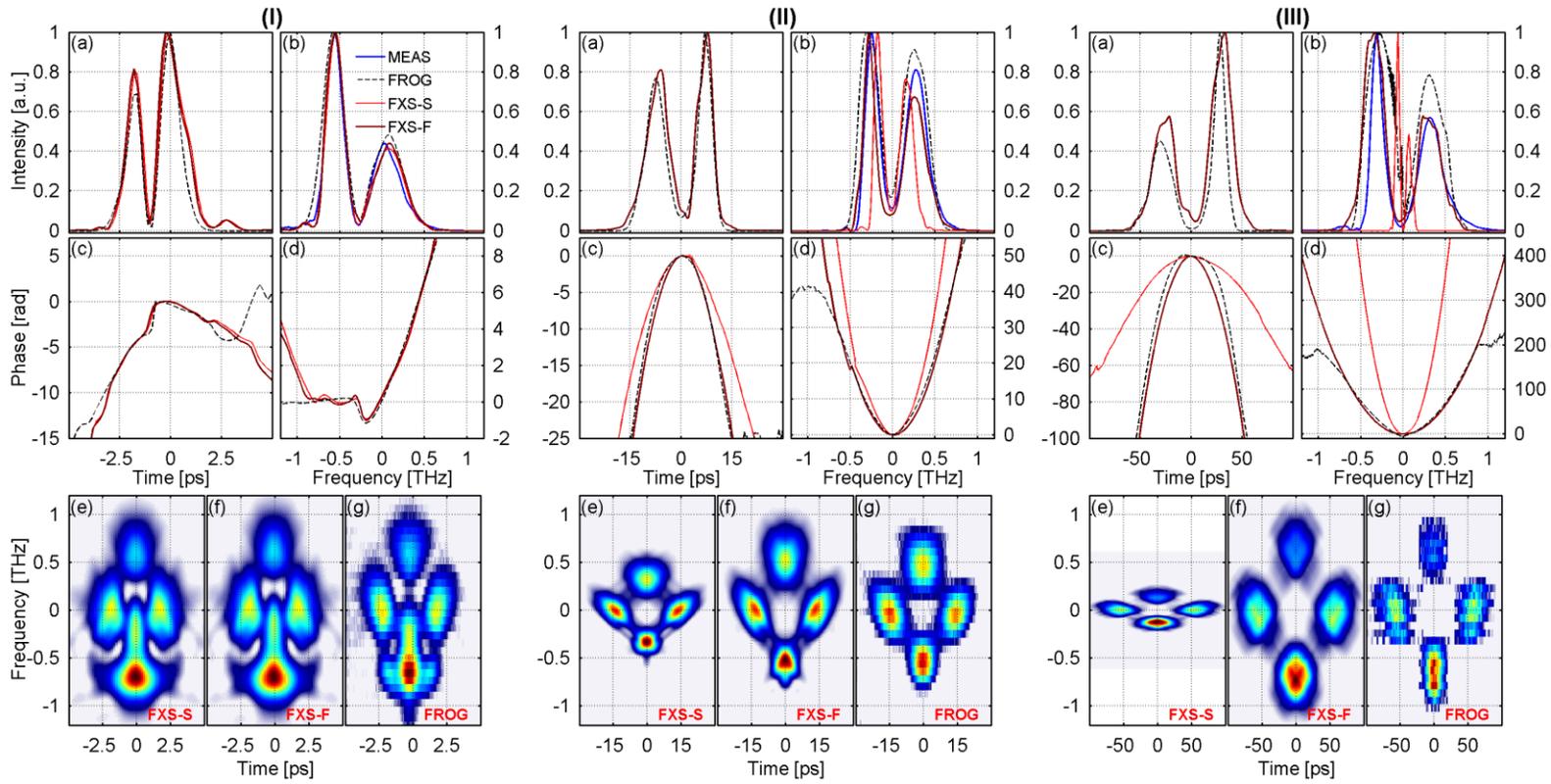

Figure 3: Measurement of optical pulses phase and amplitude performed with the FWM-X-SPIDER device, for a complex waveform of ~ 1THz bandwidth: retrieved waveform from spectra in Figure (2) and comparison with experimentally measured FROG SHG spectrograms.

Panel (I), (II) and (III) display pulses with a time-bandwidth product 5, 30 and 100 respectively. For each panel, the amplitude (a,b) and the phase (c,d) are reconstructed in the temporal and spectral domains, respectively. The red and brown lines depict the results obtained through the application of the standard (FXS-S) and novel *(Fresnel)* (FXS-F) phase-reconstruction algorithms on the measured spectral interferogram, respectively, while the dashed curves are the profiles extracted from the SHG-FROG measurement. In (b) the directly measured spectrum is also shown (in blue).

For each panel, (e)-(f) show the numerically reconstructed FROG spectrograms for the pulse profiles retrieved from the standard and the novel phase-reconstruction algorithms, respectively, while (g) reports the SHG-FROG measurements performed to characterize the same PUT.

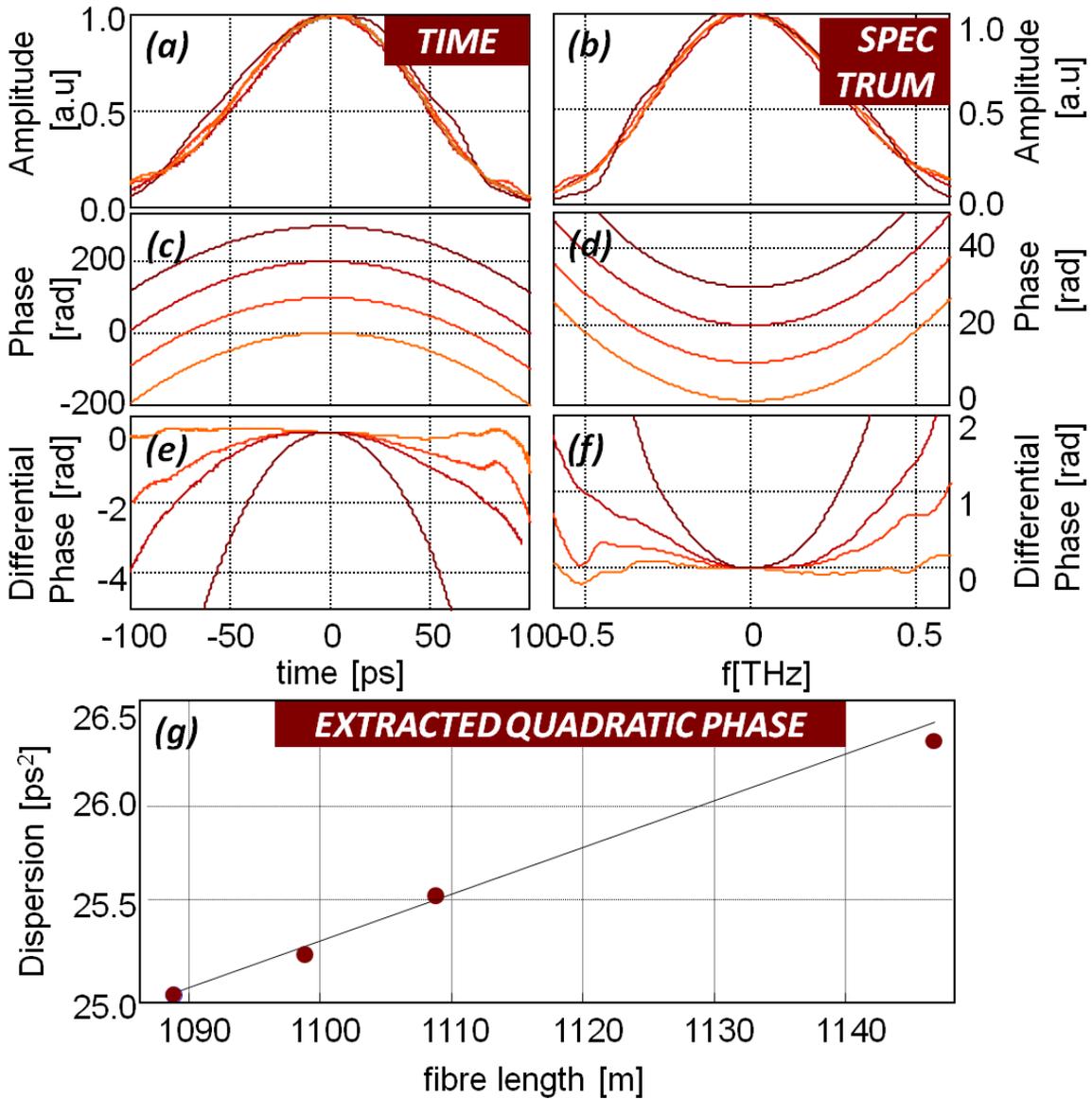

Figure 4: Reconstruction of highly-chirped Gaussian pulses, applying the novel *(Fresnel)* reconstruction algorithm. Here the pump is a 8nm bandwidth (FWHM=0.85THz) Gaussian pulse centred at 1530nm and stretched to ~150ps with 600pJ energy; its phase curvature was ~40ps$^2$, resulting in an equivalent chirp for the nonlinear interaction of 20ps$^2$ (see Methods for details).

(a-d) Amplitude and phase-reconstruction, of a 5-nm (FWHM=0.53THz) bandwidth Gaussian pulse centred at 1550nm through different SMF sections with lengths of 1090m, 1100m, 1110m and 1150m (light orange, orange, red and dark red lines, respectively), in time (a, c) and frequency (b ,d). (e-f) Differential phase: phase of the pulses in (c-d) after the subtraction of a parabola with a 25ps$^2$ curvature, equivalent to 1090m of SMF dispersion, in time and frequency, respectively. (g) The quadratic term extracted from the recovered spectral phase profiles in (d). The obtained first-order dispersion coefficients (dots) are in very good agreement with the expected values, according to the group-velocity dispersion of the SMF used in our experiments (18ps/(nm km)) (black line).